# The Physical Characterization of the Potentially-Hazardous Asteroid 2004 BL86: A Fragment of a Differentiated Asteroid

Vishnu Reddy[1]
Planetary Science Institute, 1700 East Fort Lowell Road, Tucson, AZ 85719, USA
Email: reddy@psi.edu

Bruce L. Gary
Hereford Arizona Observatory, Hereford, AZ 85615, USA

Juan A. Sanchez[1]
Planetary Science Institute, 1700 East Fort Lowell Road, Tucson, AZ 85719, USA

Driss Takir[1]
Planetary Science Institute, 1700 East Fort Lowell Road, Tucson, AZ 85719, USA

Cristina A. Thomas[1]
NASA Goddard Spaceflight Center, Greenbelt, MD 20771, USA

Paul S. Hardersen[1]
Department of Space Studies, University of North Dakota, Grand Forks, ND 58202, USA

Yenal Ogmen
Green Island Observatory, Geçitkale, Mağusa, via Mersin 10 Turkey, North Cyprus

Paul Benni
Acton Sky Portal, 3 Concetta Circle, Acton, MA 01720, USA

Thomas G. Kaye
Raemor Vista Observatory, Sierra Vista, AZ 85650

Joao Gregorio
Atalaia Group, Crow Observatory (Portalegre) Travessa da Cidreira, 2 rc D, 2645-039 Alcabideche, Portugal

Joe Garlitz
1155 Hartford St., Elgin, OR 97827, USA

David Polishook[1]
Weizmann Institute of Science, Herzl Street 234, Rehovot, 7610001, Israel

Lucille Le Corre[1]
Planetary Science Institute, 1700 East Fort Lowell Road, Tucson, AZ 85719, USA

Andreas Nathues

Max-Planck Institute for Solar System Research, Justus-von-Liebig-Weg 3, 37077 Göttingen, Germany

Pages: 27
Figures: 8
Tables: 4

**Proposed Running Head:** 2004 BL86: Fragment of Vesta

**Editorial correspondence to:**

Vishnu Reddy

Planetary Science Institute

1700 East Fort Lowell Road, Suite 106

Tucson 85719

(808) 342-8932 (voice)

reddy@psi.edu

## Abstract

The physical characterization of potentially hazardous asteroids (PHAs) is important for impact hazard assessment and evaluating mitigation options. Close flybys of PHAs provide an opportunity to study their surface photometric and spectral properties that enable identification of their source regions in the main asteroid belt. We observed PHA (357439) 2004 BL86 during a close flyby of the Earth at a distance of 1.2 million km (0.0080 AU) on January 26, 2015, with an array of ground-based telescopes to constrain its photometric and spectral properties. Lightcurve observations showed that the asteroid was a binary and subsequent radar observations confirmed the binary nature and gave a primary diameter of 300 meters and a secondary diameter of 50-100 meters. Our photometric observations were used to derive the phase curve of 2004 BL86 in the V-band. Two different photometric functions were fitted to this phase curve, the IAU H-G model (Bowell et al. 1989) and the Shevchenko model (Shevchenko 1996). From the fit of the H-G function we obtained an absolute magnitude H=19.51±0.02 and a slope parameter G=0.34±0.02. The Shevchenko function yielded an absolute magnitude of H=19.03±0.07 and a phase coefficient $b$=0.0225±0.0006. The phase coefficient was used to calculate the geometric albedo (Ag) using the relationship found by Belskaya and Schevchenko (2000), obtaining a value of Ag=40±8% in the V-band. With the geometric albedo and the absolute magnitudes derived from the H-G and the Shevchenko functions we calculated the diameter (D) of 2004 BL86, obtaining D=263±26, and D=328±35 meters, respectively. 2004 BL86 spectral band parameters and pyroxene chemistry are consistent with non-cumulate eucrite meteorites. A majority of these meteorites are derived from Vesta and are analogous with surface lava flows on a differentiated parent body. A non-diagnostic spectral curve match using Modeling for Asteroids tool yielded a best match with non-cumulate eucrite Bereba. Three other NEAs (1993 VW, 1998 KK17 and 2000 XH44) that were observed by Burbine et al. (2009) also have spectral properties similar to 2004 BL86. The presence of eucrites with anomalous oxygen isotope ratios compared to the howardites, eucrites, and diogenites meteorites from Vesta suggests the possible presence of multiple differentiated bodies in the inner main belt or the contamination of Vesta's surface with exogenic material. The spectral properties of both anomalous and Vestan eucrites are degenerate making it difficult to identify the parent bodies of anomalous eucrites in the main belt and the NEO population using remote sensing. This makes it difficult to link 2004 BL86 directly to Vesta, although the Vesta family is the largest contributor of V-types to near-Earth space.

## 1. Introduction

Close flybys of near-Earth asteroids (NEAs) are interesting because they offer a rare opportunity to physically characterize their surface properties with small ground-based telescopes and help enhance our knowledge about the hazards they pose and the source regions from which they are derived. The potentially hazardous asteroid (PHA) (357439) 2004 BL86 made a close flyby of the Earth at a distance of 1.2 million km (0.0080 AU) on January 26, 2015. The asteroid was discovered by the LINEAR survey on January 30, 2004. During its close flyby the asteroid's visual magnitude peaked at 9 making it easy to characterize for small ground-based telescopes. Based on lightcurve observations of 2004 BL86 Pollock et al. (2015) suggested that this asteroid is a binary. Goldstone delay-Doppler radar images confirmed its binary nature and gave a primary diameter of 300 meters and a secondary diameter of 50-100 meters (Pollock et al. 2015).

We observed 2004 BL86 during its close flyby to constrain its physical properties. Our primary goal was to constrain its surface composition and identify a meteorite analog for 2004 BL86. A secondary goal was to derive its photometric phase function given the rapid change in phase angle during the close flyby.

This paper is divided into two main sections, the photometric study presented in section 2, and the spectroscopic study discussed in section 3. The photometric study includes a description of the observations and data reduction procedure, and the analysis of the phase curve of 2004 BL86. The spectroscopic study, on the other hand, focuses on the mineralogical analysis of the asteroid and its relationship with meteorites and other objects that exhibit the same spectral features. In section 4 we list our main findings.

## 2. Photometric Study

### 2.1 Observations and Data Reduction

2004 BL86 was observed during a 7-day interval starting 0.5 days before its closest approach (January 27.18, 2015 UTC). The observers and observing circumstances are listed in Table 1. During the seven days of observations, nine rotation light curves were obtained. Due to the asteroid's fast motion across the sky during the dates close to opposition, it was necessary for each observer to change their field of view (FOV) positions several times during observing sessions. For example, observer Y. Ogmen used 37 FOV settings for the January 26 4.6 hr observing session. This required the use of 37 different sets of calibration stars, typically 4 per FOV. It was important to check for internal consistency between all calibration stars for each FOV in order to identify variables and exclude their use. Since 2004 BL86 exhibited a peak-to-peak rotation variation of $\sim$ 0.10 mag it was important to calibrate each FOV with an accuracy of less than $\sim$ 0.015 mag. This was achieved for almost all FOVs for all observers.

Standard differential photometry procedures were employed for producing light curves. The number of calibration stars used per FOV ranged from 2 to 30, though 4 was typical. Most calibrations included "star color sensitivity" corrections (serving the same purpose as the classical "CCD transformation equations"). AAVSO Photometric All-Sky Survey (APASS) V- and r'-magnitudes were used for calibration; the choice of which magnitude system to use was dictated by which one afforded the smallest "star color sensitivity." For example, clear filter observations were typically calibrated using r' magnitudes, though in one case V-magnitudes were preferred.

When an asteroid's phase angle varies slowly, and when its distance from Earth is also varying slowly (on a percentage basis), there is a straightforward process for creating a rotation lightcurve: time is converted to a rotation phase and magnitudes are plotted vs. this phase. For all but the last of the observing sessions reported here this could not be done, because both phase angle and distance from Earth varied significantly during an observing session. A procedure was employed for fitting measured magnitude vs. time (JD) that included the following three components of variation: 1) changing distance from Earth, 2) changing phase angle (i.e., the HG phase effect function), and 3) variation due to rotation (including several harmonics of the variation with the shortest period: P = ½ rotation period). After such a solution it is possible to isolate each component and produce rotation lightcurves and a phase curve plot.

Figure 1 shows six of the highest quality rotation lightcurves, from before opposition to six days afterward. Vertical offsets are used for clarity. The rotation period adopted for phase-folding is 2.620 hours. This period was determined from lightcurve data with a precision that is much less than reported by Pollock et al. (2015), whose value of 2.6205 ± 0.0005 hours was based on many observations from Chile made prior to closest approach. Observations made during closest approach are unsuitable for determining rotation period because that's when rotation lightcurves shape is expected to change the most due to changing viewing geometry (phase angle and distance to Earth). Indeed, although our phase-folded lightcurves exhibited gross properties that remained the same, new features appeared at the time of closest approach and are superimposed upon the gross shape. All phase-folded rotation lightcurves show a minimum at phase ∼ 0.35. Most exhibit a maximum at phase ∼ 0.05. After closest approach a pair of maxima appear at ∼ 0.55 and 0.80, replacing what was a broad bright interval (0.6-1.0) seen in the first lightcurve (by Ogmen).

The Spectral Energy Distribution (SED) of 2004 BL86 was obtained in order to determine star colors (e.g., B-V, g'-r', V-r', V-Rc) for use with the star color sensitivity calibrations mentioned above. These colors are also used for converting calibrated magnitudes (such as r') to V, which was necessary for deriving a phase effect solution.

The SED was obtained in two ways: 1) using g'r'i' filters and corresponding APASS magnitudes for background calibration stars, and 2) Star Analyzer 100 (SA-100) transmission grating measurements calibrated using three solar analog stars. Both measurements were made on January 27.

The g'r'i' measurements were made in the traditional way, using differential photometry and 20 to 30 stars for calibration. Star color sensitivity concepts were employed, as described above.

The visible spectrum measurements made with the g'r'i' filters consist of 3 magnitudes for a specific date and UT (2015.01.27, 05.4 UT). These were converted to fluxes using the following relation:

$$\lambda F_\lambda \ [W/m^2] = \text{SED Constant} \times 10^{-0.4 \, \text{Mag}} \tag{1}$$

where the "SED Constants" for g', r' and i' are 2.48e-8, 1.75e-8 and 1.40e-8, and "Mag" corresponds to g'-mag, r'-mag and i'-mag.

The SA-100 transmission grating was mounted to a 0.35-m F/1.9 telescope. The spectral resolution with this grating setup is ~100, similar to the SpeX instrument on the NASA IRTF (Rayner et al. 2005). Measurements were made on January 27, from 07:00 to 08:15 UTC when the asteroid was at V-mag 9.8 and phase angle 5°. Spectral calibration was accomplished using solar analog stars SAO 80423, SAO 98168 and SAO 43046. The observing sequence was the usual one for minimizing atmospheric extinction trends: star1, 2004 BL86, star2, 2004 BL86, star3. The air mass values were all < 1.05.

The SA-100 is a diffraction grating with a groove spacing of 100 lines per millimeter, and produces images, including a zero-order star-like PSF (16% of incident light), a first-order spectrum (78% of incident light) and a second-order spectrum with non-overlap of the first-order spectrum permitting measurements between 0.4 μm and 1.0 μm. First-order spectra in each of 22 10-second exposure images of 2004 BL86 were aligned (using the zero-order image), and these were median-combined in order to remove the unwanted registration of background stars.

The SA-100 spectrum was obtained 1.85 hours after the g'r'i' observations. The three calibration solar analog stars used to calibrate the 2004 BL86 spectrum were assigned magnitudes based on Bt/Vt catalog values (Tycho-2 B and V magnitudes, converted to standard B and V magnitudes). APASS magnitudes were not available for them because they were too bright to be included in that catalog. Solar analog spectra were multiplied by the V-band spectral response function, as a weighting function, to obtain an equivalence between SA-100 ADU counts pixel$^{-1}$ and flux (W m$^{-2}$). The solar analog fluxes, weighted by the V-band response function, were obtained using the same equation above, relating $\lambda F_\lambda$ to magnitude (where the SED Constant for V-band is 2.09e-8).

In order to compare the three g'r'i' SED values with the SA-100 spectrum it was necessary to apply an offset of 0.27 ± 0.01 magnitude. This offset was applied due to changes in brightness as the Earth/asteroid distance and phase angle changed during the flyby. In order to determine 2004 BL86's colors for bands other than g'r'i', "pseudo" magnitudes were found by trial and error that matched the visible wavelength spectrum. In this way we determined that B-V=+0.95, g'-r'=+0.607, V-r'=+0.276, V-g'=-0.331, and V-Rc=+0.5.

2.2 Phase Curve Analysis

In creating a plot of brightness versus phase angle it is customary to adopt a rotation phase corresponding to maximum brightness, since minima vary more than maxima. The phase angle (sun-target-Earth) varied during our observations from 1.5° to 54°. Four of the eleven observing sessions were made when the phase angle was less than 7°, so our data sampled the "opposition effect" (seen in Fig 2).

Figure 2 shows the phase curve with all measurements converted to V-band. When measurements were made with an r'-band filter, for example, the images were processed in a way that produced r'-magnitudes on the assumption that 2004 BL86 had a g'-r' color of +0.607 (derived in the previous section). Images made with a R-band filter were processed in a way that produced r'-magnitudes using the same BL86 color assumption. Clear filter images (and those with a "clear with blue-blocking" filter) were processed the same way, resulting in r'-magnitudes. The G filter observations (using a Santa Barbara Instrument Group G band filter) were processed as if a V-band filter was used, assuming 2004 BL86 had a g'-r' color of +0.607. Converting r'-magnitudes to a V-magnitude equivalent was achieved by adding 0.276. In other words, all images were processed using a process that was equivalent to applying "CCD transformation equations."

The phase curve plot uses labels meant to indicate which filter was used for the observations and which band was used for calibration. For example, C/V means that a clear filter (or "clear with blue blocking") was used for the observations and APASS V-band magnitudes were used for calibration. G/V means that a G filter was used and APASS V magnitudes were used for calibration. The "Ogmen V/V" data were V/V observed before closest approach.

Changes in viewing direction during the week of phase curve observations can distort the phase curve in ways that depend upon the orientation of the spin axis. We lack knowledge of the rotation axis orientation so this effect cannot be modeled.

According to radar observations the ratio of solid angles of the secondary to primary is ~6% (3%-12%). Since no mutual events were observed we assume that our photometric measurements are for the sum of the optical fluxes of both bodies.

We fitted two different photometric functions to the phase curve, the IAU H-G model (Bowell et al. 1989) and the Shevchenko model (Shevchenko 1996). For these models we developed a Python routine that makes use of the curve_fit function included in the SciPy library for Python. This function uses the Levenberg-Marquardt algorithm, which performs a non-linear least squares fit of the function to the data. The routine returns the optimal values for the parameters so that the sum of the squares of the differences between the function and the data ($\chi^2$) is minimized. The uncertainties in the fit parameters are given by the 1-sigma calculated from the covariance matrix.

The IAU H-G model (Bowell et al. 1989) is described by the following two-parameter function:

$$H(\alpha) = H - 2.5\log[(1-G)\phi_1(\alpha) + G\phi_2(\alpha)] \tag{2}$$

where $H(\alpha)$ is the reduced magnitude, $\alpha$ is the phase angle, $H$ is the absolute magnitude, $G$ is the slope parameter, which describes the shape of the phase curve, and $\phi_1$ and $\phi_2$ are phase functions.

From our fit we obtained $H = 19.51 \pm 0.02$ and $G = 0.34 \pm 0.02$. Figure 2 shows the phase curve obtained using the IAU H-G function (red solid line). As can be seen in this figure the function tends to deviate from the data at the lowest phase angles.

The Shevchenko model (Shevchenko 1996) is a three-parameter function with the following form:

$$H(\alpha) = H + [a\alpha/(1+\alpha)] + b\alpha \tag{3}$$

where $a$ characterizes the amplitude of the opposition effect, and the phase coefficient $b$ describes the linear part of the phase curve. We obtained values of $H = 19.03 \pm 0.07$, $a = 0.86 \pm 0.09$ and $b = 0.0225 \pm 0.0006$. The phase curve obtained with the Shevchenko function is depicted in Figure 2 as a dashed line. This function seems to provide a better fit to the data obtained at low phase angles, close to the opposition surge.

Belskaya and Schevchenko (2000) have shown that the database of large asteroids exhibits a relationship between the geometric albedo (Ag) and the phase coefficient ($b$). The suggested relationship is:

$$b = 0.013(\pm 0.002) - 0.024(\pm 0.002) \times \text{Log (Ag)} \tag{4}$$

This equation can be used to solve for the albedo by knowing the phase coefficient. Substituting $b = 0.0225 \pm 0.0006$ and solving for the albedo yields Ag = $0.40 \pm 0.08$ at V-band ($\sim 0.54$ $\mu$m).

The bolometric Bond albedo ($A_B$) can be linked to the geometric albedo through:

$$A_B = qA_g \qquad (5)$$

where q is the phase integral and can be calculated as

$$q = 0.290 + 0.684 \ast G \qquad (6)$$

given a value of q=0.525±0.011. Thus the bolometric Bond albedo is $A_B$ = 0.21±0.04.

The absolute magnitude and the geometric albedo can be used to estimate the diameter of the asteroid using the following equation (Fowler and Chillemi, 1992; Pravec and Harris , 2007):

$$D = [1329/(A_g)^{1/2}] \times 10^{-H/5} \qquad (7)$$

We found that for H=19.51±0.02 (IAU H-G), D=263±26 m, and for H=19.03±0.07 (Shevchenko model), D=328±35 m. The IAU H-G function is known for its limitation when modeling the opposition effect (e.g., Belskaya and Schevchenko 2000; Hasegawa et al. 2014; Ishiguro et al. 2014), therefore the values derived for the absolute magnitude and diameter of 2004 BL86 using the Shevchenko function are probably more accurate than those derived using the H-G function. We also notice that, within the uncertainty, the diameter of the asteroid calculated using the H value from the Shevchenko function is more consistent with the diameter derived from radar observation ($\sim$ 300 m). The parameters derived from the two photometric functions are presented in Table 2.

## 3. Spectroscopic Study

### 3.1 Observations and Data Reduction

Near-infrared (IR) spectroscopic observations of 2004 BL86 were obtained remotely using the SpeX instrument of the NASA Infrared Telescope Facility (IRTF) in prism mode (Rayner et al. 2003) on January 26, 2015 between 9:14 UTC and 12:13 UTC when the asteroid was V-mag 10.25. Apart from the asteroid, local G-type star SAO174759 and solar analog star SAO120107 were also observed for telluric and solar slope corrections. The observational circumstances for the near-IR observations are given in Table 3.

The processing of the prism data was carried out using the IDL-based Spextool provided by the NASA IRTF (Cushing et al., 2004). Analysis of the data to determine spectral band parameters like band centers, band depths and Band Area Ratio (BAR) was performed using a Matlab code based on the protocols discussed by Cloutis et al. (1986). Spectral band parameters were measured for two sets of data and were averaged before mineralogical analysis. The errors of the band parameters are given by the 1-σ (standard deviation of the mean) estimated from multiple measurements of each band parameter.

Spectral band parameters such as band centers and band areas can be affected by a change in temperature (e.g., Reddy et al., 2012a; Sanchez et al., 2012). Therefore, we used the equation from Burbine et al. (2009) to calculate the surface temperature of 2004 BL86 at the time of observations, and equations from Reddy et al. (2012a) to correct for temperature effects and normalize the data to room temperature. Laboratory spectral calibrations from Burbine et al. (2007) were used for the mineralogical analysis. Spectral band parameters for 2004 BL86 are given in Table 4.

3.2 Surface Composition and meteorite analogs

Figure 3 shows the near-IR spectrum of 2004 BL86 with deep absorption bands due to pyroxene. Figure 4 shows the Band I vs. Band II centers of clino and orthopyroxenes along with howardites, eucrites and diogenites (HED) meteorites from Vesta. Howardites, eucrites and diogenites differ from each other primarily based on their pyroxene composition and pyroxene to plagioclase ratio. These ratios were estimated to be 3:1 for howardites, 1:1 for eucrites, and 30:1 for diogenites (Gaffey, 1976; Mittlefehldt et al., 1998). The Fe content in low-Ca pyroxene in HEDs increases from diogenites ($Fs_{20-33}$), to cumulate eucrites ($Fs_{30-44}$), to basaltic eucrites ($Fs_{43-55}$). Eucrites are basaltic melts from partial melting and can be surface lava flows (non-cumulate eucrites) or subsurface cumulate layers (cumulate eucrites) formed by fractional crystallization. Diogenites are orthopyroxenites that formed at great depths (lower crust/upper mantle) and cooled slowly. Howardites are physical mixtures of eucrites and diogenites (regolith breccias). The pyroxene chemistry of cumulate eucrites overlaps that of howardites. 2004 BL86 plots closer to the eucrites than howardites and diogenites in the Band I vs. Band II plot (Fig. 4). Similarly, in the Band I center vs. BAR plot (Fig. 5), 2004 BL86 is located in the basaltic achondrites region, closer to eucrites than diogenites, confirming its igneous origin.

A non-diagnostic spectral curve match search using the Modeling for Asteroids tool (M4AST) (Popescu et al. 2012) yielded a best match with monomict eucrite Bereba (Fig. 6). Mineralogical analysis was performed using equations from Burbine et al. (2007) where the Band I and II centers were used to calculate the molar ferrosilite (Fs) and wollastonite (Wo) content of the pyroxene. 2004 BL86 has a mean pyroxene composition of $Fs_{45.2\pm3}$ and $Wo_{10.7\pm1}$. These values are consistent with non-cumulate eucrites, as shown in Figure 7.

3.3 Relationship with Vesta, Vestoids and V-types.

The mineralogy and spectral band parameters of 2004 BL86 are consistent with HED meteorites. A majority of these meteorites are derived from Vesta and Vestoids, which are fragments of Vesta excavated during the formation of the two large basins on its south pole (Russell et al. 2012). On a hemispherical scale, Vesta's surface is consistent with howardites-like composition (Reddy et al. 2012b; De Sanctis et al. 2013). Dawn data have not identified any distinct lithologic units on Vesta that are pure eucrites or diogenites although howardites that are rich in diogenites dominate the southern hemisphere of Vesta (Russell et al. 2012). Vesta's absorption bands are also subdued compared to 2004 BL86 (Fig. 8). This could be due to the presence of hemispherical scale dark carbonaceous chondrite impactor material on Vesta that reduces the albedo and band depth of pyroxene absorption bands (Reddy et al. 2012c).

We compared the pyroxene chemistry of 2004 BL86 with V-type NEAs observed by Burbine et al. (2009). As can be seen in Figure 7, 2004 BL86 and the three NEAs (1993 VW, 1998 KK17 and 2000 XH44) fall in the non-cumulate eucrites (surface flows) zone of the plot, suggesting similar surface compositions for all of these objects. While the composition and spectral band parameters of 2004 BL86 are consistent with non-cumulate eucrites, not all non-cumulate eucrites in the meteorite collection come from Vesta. Six anomalous eucrites (NWA 1240; Pasamonte; A-881394; Caldera; Bunburra Rockhole; and Ibitira) all show oxygen isotope ratios that are distinct from the HEDs (Wiechert et al., 2004; and Greenwood et al., 2005; Bland et al., 2009). The implications of these anomalous oxygen isotope ratios are: a) the parent body of the HED meteorites did not differentiate entirely and remained isotopically heterogeneous (Wiechert et al., 2004); b) the anomalous oxygen isotopes of the HED meteorites are due to contamination of exogenic impactors (Greenwood et al., 2005), and c) there are multiple parent bodies for HED meteorites in the inner main belt as evidenced by the Aten-type orbit of the Bunburra Rockhole meteorite (Bland et al., 2009).

Data from the Dawn mission provide convincing evidence that Vesta differentiated with a core/mantle/crust early in our solar system history (Russell et al. 2012). This leaves options b and c to explain the anomalous non-cumulate eucrites seen in the HED collection. Laboratory analysis of HED meteorites and data from the Dawn mission have shown extensive contamination of Vesta's surface by exogenic meteorites. Wee et al. (2010) analyzed 13 howardites and four polymict eucrites to document exogenic materials in HED breccia. Their analysis indicates possible contamination of the Vestan surface with a variety of chondritic materials including CI, CK, CM, EH, EL, H, L and LL chondrites. Reddy et al. (2012b) and Le Corre et al. (2015) discovered evidence for carbonaceous chondrite and ordinary chondrite impactors on Vesta from Dawn data. Hence, it is possible that anomalous oxygen isotopes of some eucrites could be due to a contamination from exogenic material. The Aten-type orbit of the Bunburra Rockhole meteorite suggests that it was derived from the inner asteroid belt with a 72% probability that it was delivered via v6 resonance and only a 2% chance that it was delivered from the outer main belt via 3:1 resonance (Bland et al. 2009). Spectral properties of anomalous eucrites are not different from those of Vestan eucrites and it is challenging to distinguish V-type asteroids from non-Vesta parent bodies. Based on these factors, it is not possible to directly link 2004 BL86 to Vesta or another differentiated parent body in the inner main belt, although the Vesta family is the largest contributor of V-types to near-Earth space.

## 4. Summary

We observed binary near-Earth asteroid (357439) 2004 BL86 during a close flyby of the Earth at a distance of 1.2 million km (0.0080 AU) on January 26, 2015. Our comprehensive photometric and mineralogical analysis sheds light on the objects surface properties and meteorite affinities and source region from which it was derived from. Based on our study here is what we determined:

- From the fit of the H-G function to the phase curve we obtained values of H=19.51±0.02 and G=0.34±0.02. The use of the Shevchenko function yielded an absolute magnitude of H=19.03±0.07 and a phase coefficient $b$=0.0225±0.0006.

- The geometric albedo (Ag) of 2004 BL86 in the V-band is 40±8%. Using this value and the absolute magnitudes derived from the H-G function and the Shevchenko function we calculated the diameter of 2004 BL86, obtaining values of D=263±26 m, and D=328±35 m, respectively.

- The 2004 BL86 spectral band parameters and pyroxene chemistry are consistent with non-cumulate eucrite meteorites. A majority of these meteorites are derived from Vesta and are analogous to surface lava flows on a differentiated parent body.

- A non-diagnostic spectral curve match using Modeling for Asteroids tool yielded a best match with the non-cumulate eucrite Bereba.

- Three other NEAs (1993 VW, 1998 KK17 and 2000 XH44) observed by Burbine et al. (2009) also have spectral properties similar to 2004 BL86.

- Spectral properties of both anomalous and Vestan eucrites are degenerate, making it difficult to identify parent bodies of anomalous eucrites in the main belt and the NEO population using remote sensing. This makes it difficult to link 2004 BL86 directly to Vesta, although the Vesta family is the largest contributor of V-types to near-Earth space.

## Acknowledgments

This research work was supported by the NASA Near-Earth Object Observations Program grant NNX14AL06G (PI:Reddy). We thank the IRTF TAC for awarding time to this project, and to the IRTF TOs and MKSS staff for their support. The IRTF is operated by the University of Hawaii under Cooperative Agreement no. NCC 5-538 with the National Aeronautics and Space Administration, Office of Space Science, Planetary Astronomy Program.

**Tables**

**Table 1:** Observers and observational circumstances for photometric observations. Geocentric range is given for the start UTC.

| Date UTC | Observer and Telescope | Filter | UT range | Phase Angle (°) | Geocentric Range (AU) | Magnitude |
|---|---|---|---|---|---|---|
| Jan. 26, 2015 | Y. Ogmen, 14" Schmidt-Cassegrain | V | 18.2-22.8 | 25.5-12.7 | 0.0080 | 10.0-9.7 |
| Jan. 26, 2015 | J. Gregorio, 12" Schmidt-Cassegrain | V | 23.0-24.8 | 12.7-08.0 | 0.0084 | 9.7-9.6 |
| Jan. 27, 2015 | B. Gary, 14" Schmidt-Cassegrain | g'r'i' | 05.2-10.0 | 02.0-09.6 | 0.0093 | 9.1-9.7 |
| Jan. 27, 2015 | T. Kaye, 5" refractor | Cb | 05.9-10.2 | 03.0-09.8 | 0.010 | 9.2-9.7 |
| Jan. 27, 2015 | J. Garlitz, 12" Newtonian | G | 04.2-07.5 | 01.5-05.5 | 0.0091 | 9.0-9.4 |
| Jan. 28, 2015 | T. Kaye, 5" refractor | Cb | 04.5-06.0 | 28.9-30.2 | 0.016 | 11.2-11.4 |
| Jan. 29, 2015 | P. Benni, 11" Schmidt-Cassegrain | R | 00.4-11.1 | 38.2-41.3 | 0.022 | 12.3-12.7 |
| Feb. 02, 2015 | B. Gary, 14" Schmidt-Cassegrain | C | 03.8-09.3 | 49.5-49.6 | 0.059 | 14.7-14.9 |

**Table 2:** Parameters derived from the IAU H-G and Shevchenko functions. The columns in this table correspond to: absolute magnitude (H), slope parameter (G), parameters $a$ and $b$ from the Shevchenko function (see text for explanation), phase integral (q), geometric albedo ($A_g$), bolometric albedo ($A_B$), diameter (D), and $\chi^2$.

| Function | H (mag) | G | a | b (mag/°) | q | Ag | $A_B$ | D (m) | $\chi^2$ |
|---|---|---|---|---|---|---|---|---|---|
| IAU-HG | 19.51± 0.02 | 0.34± 0.02 | --- | --- | 0.525 ±0.011 | --- | 0.21± 0.04 | 263±26 | 0.035 |
| Shevchenko | 19.03± 0.07 | --- | 0.86± 0.09 | 0.0225± 0.0006 | --- | 0.40 ± 0.08 | 0.21± 0.04 | 328±35 | 0.023 |

**Table 3.** Observational circumstances for the near-IR observations.

| Date UTC | Observer and Telescope | Wavelength | UT range | Phase Angle | Mag. |
|---|---|---|---|---|---|
| Jan. 26, 2015 | V. Reddy, C. Thomas, D. Takir, D. Polishook NASA IRTF | 0.7-2.5 μm | 09.23-12.21 | 47.7-40.1 | 10.7-10.4 |

**Table 4**: Spectral band parameters and pyroxene chemistry for 2004 BL86. Data presented in this table correspond to: Band I center (BIC), temperature-corrected Band I center (ΔBIC), Band II center (BIIC), temperature-corrected Band II center (ΔBIIC), Band I depth (BID), Band II depth (BIID), Band Area Ratio (BAR), molar contents of wollastonite (Wo) and ferrosilite (Fs) with their temperature-corrected values ΔWo, and ΔFs. The average surface temperature of 2004 BL86 was calculated as in Burbine et al. (2009), temperature corrections are from Reddy et al. (2012a).

| Spectral Band Parameter | Value |
|---|---|
| BIC (μm) | 0.936±0.005 |
| ΔBIC (μm) | 0.937±0.005 |
| BIIC (μm) | 1.99±0.01 |
| ΔBIIC (μm) | 1.99±0.01 |
| BID (%) | 42.2±0.1 |
| BIID (%) | 30.8±0.2 |
| BAR | 1.89±0.04 |
| Wo (mol %) | 10.3±1.0 |
| ΔWo (mol %) | 10.7±1.0 |
| Fs (mol %) | 44.3±3.0 |
| ΔFs (mol %) | 45.2±3.0 |

**Figure Captions**

**Figure 1.** Six of the highest quality rotation light curves from before opposition to six days after opposition. The data are offset vertically for clarity. The rotation period adopted for phase-folding is 2.620 hours as reported by Pollock et al. (2015).

**Figure 2.** Phase curve of 2004 BL86. All filter bands were converted to V-band using spectral energy distribution information. Data labels convey information about what filter was used and what standard magnitude band they were calibrated for. Phase curves obtained with the IAU H-G and Shevchenko functions are depicted as a red solid line and as a dashed line, respectively.

**Figure 3.** Near-infrared spectrum of asteroid 2004 BL86.

**Figure 4.** Band I center vs. Band II center for 2004 BL86 and HED meteorites. The values for HED meteorites are from Le Corre et al. (2015). Also shown are the measured band centers for orthopyroxenes (open triangles) from Adams (1974), and clinopyroxenes (open circles) from Cloutis and Gaffey (1991).

**Figure 5.** Band I center vs. BAR for 2004 BL86 and HED meteorites. Values for HED meteorites are from Le Corre et al. (2015). The rectangular zone corresponding to the S(I)-type asteroids encompasses monomineralic olivine assemblages (Gaffey et al., 1993). The polygonal region, corresponding to the S(IV) subgroup, represents the mafic silicate components of ordinary chondrites (OC). The dashed curve indicates the location of the olivine-orthopyroxene mixing line (Cloutis et al.,1986).

**Figure 6.** Near-IR spectrum of asteroid 2004 BL86 along with the spectrum of Eucrite Bereba. The match was found using M4AST (Popescu et al. 2012).

**Figure 7.** Molar contents of Wo vs. Fs for 2004 BL86 (red diamond). The error bars correspond to the values determined by Burbine et al. (2007): 3 mol % for Fs and 1 mol % for Wo. The approximated range of pyroxene chemistries for howardites, non-cumulate eucrites, cumulate eucrites, and diogenites from Mittlefehldt et al. (1998) are indicated as dashed-line boxes. Also shown are pyroxene chemistries for three V-type NEAs (1993 VW, 1998 KK17 and 2000 XH44) observed by Burbine et al. (2009).

**Figure 8.** Near-IR spectrum of asteroid 2004 BL86 along with the spectrum of (4) Vesta. Spectra are normalized to unity at 0.75 μm.

Figure 1.

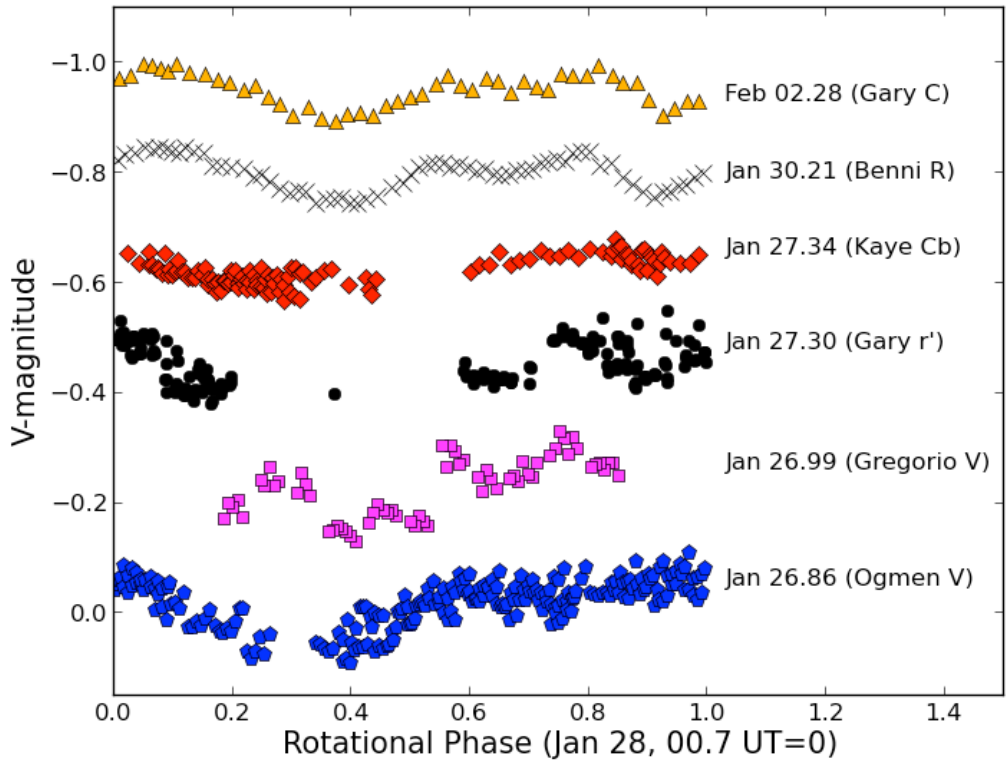

Figure 2

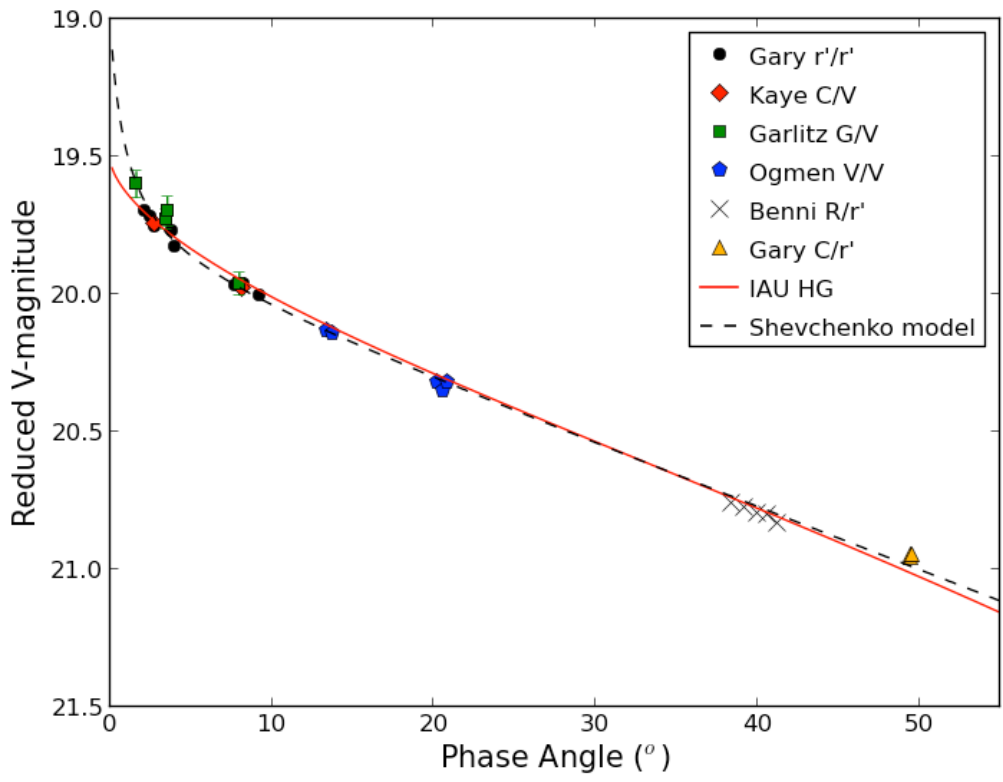

Figure 3

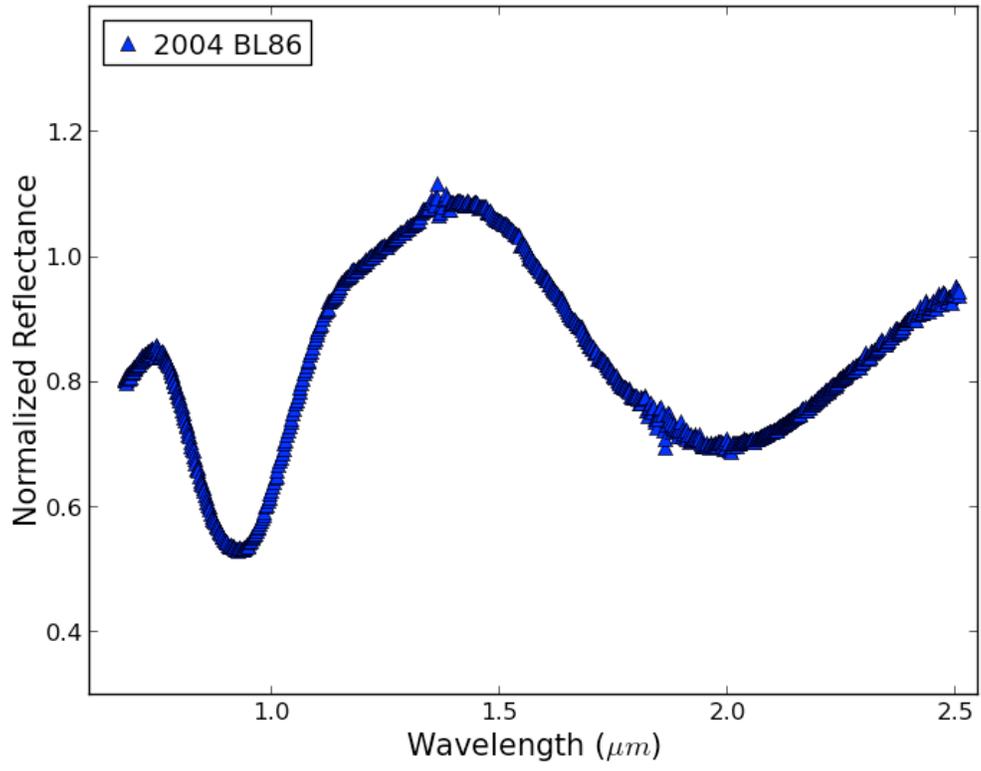

Figure 4

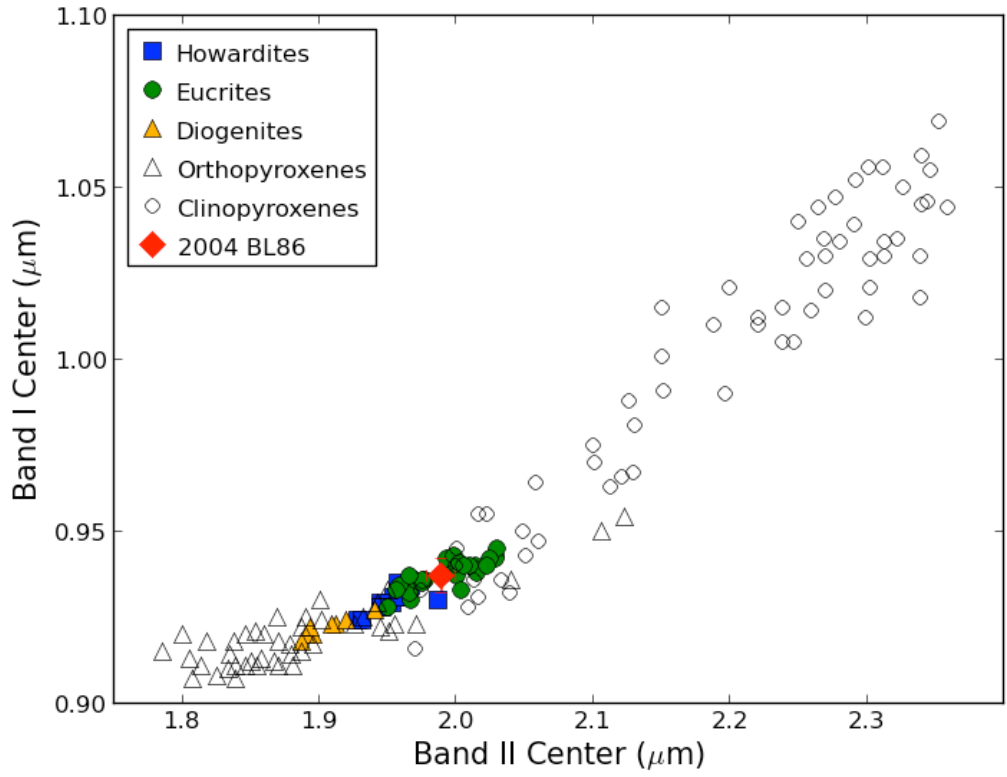

Figure 5

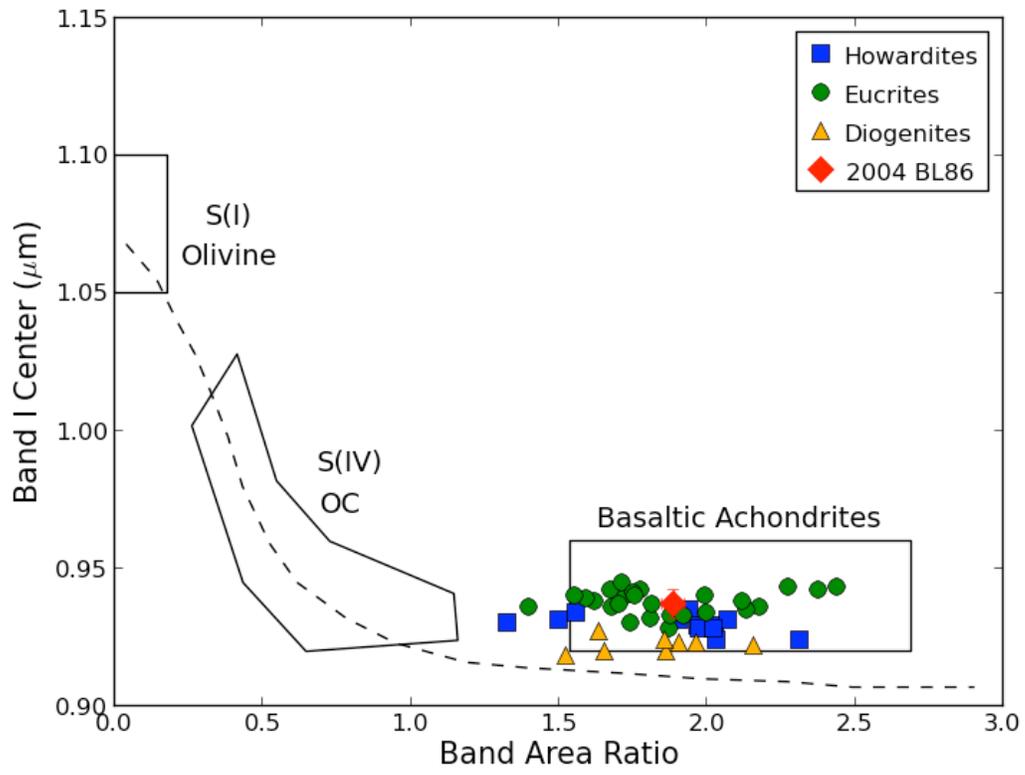

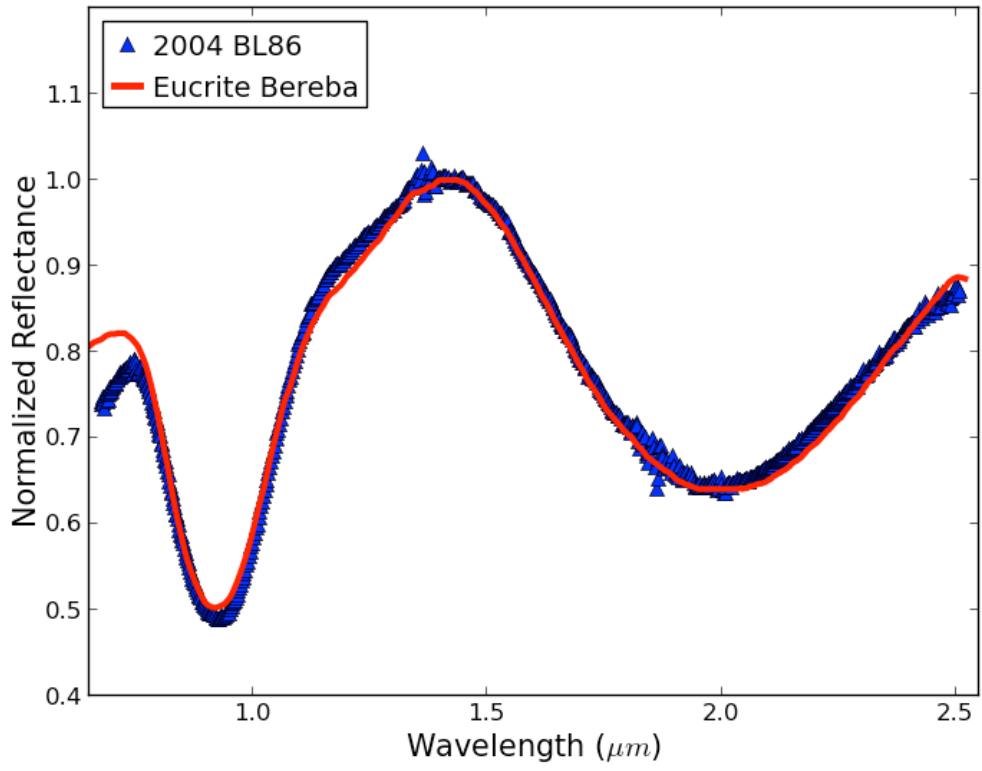

Figure 7

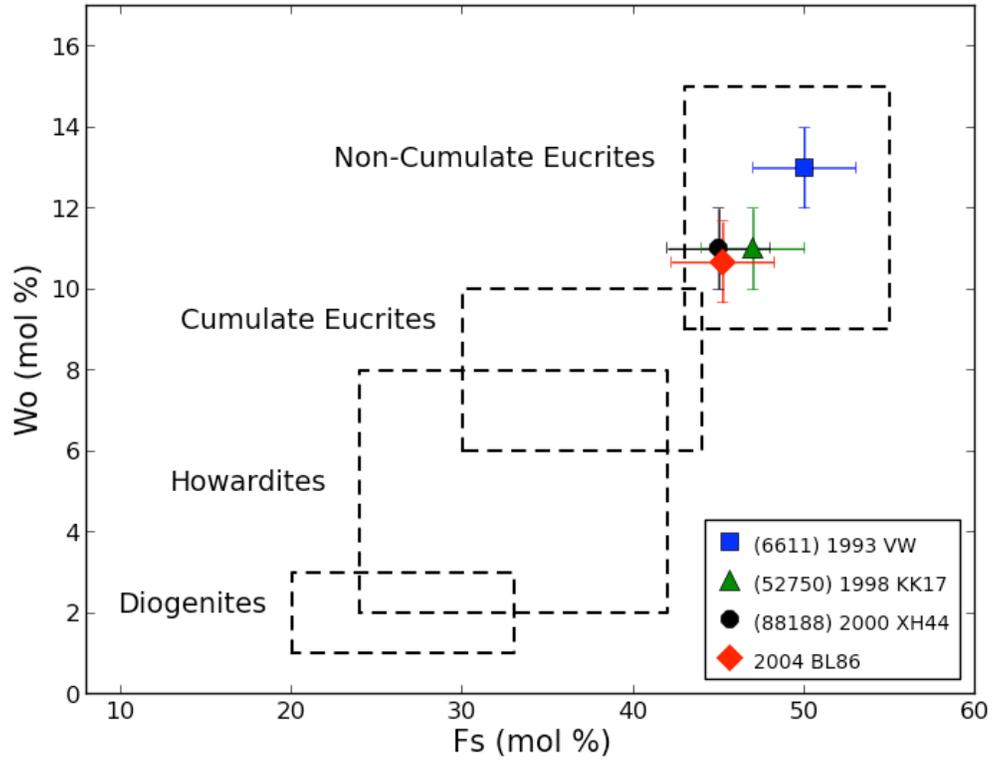

Figure 8

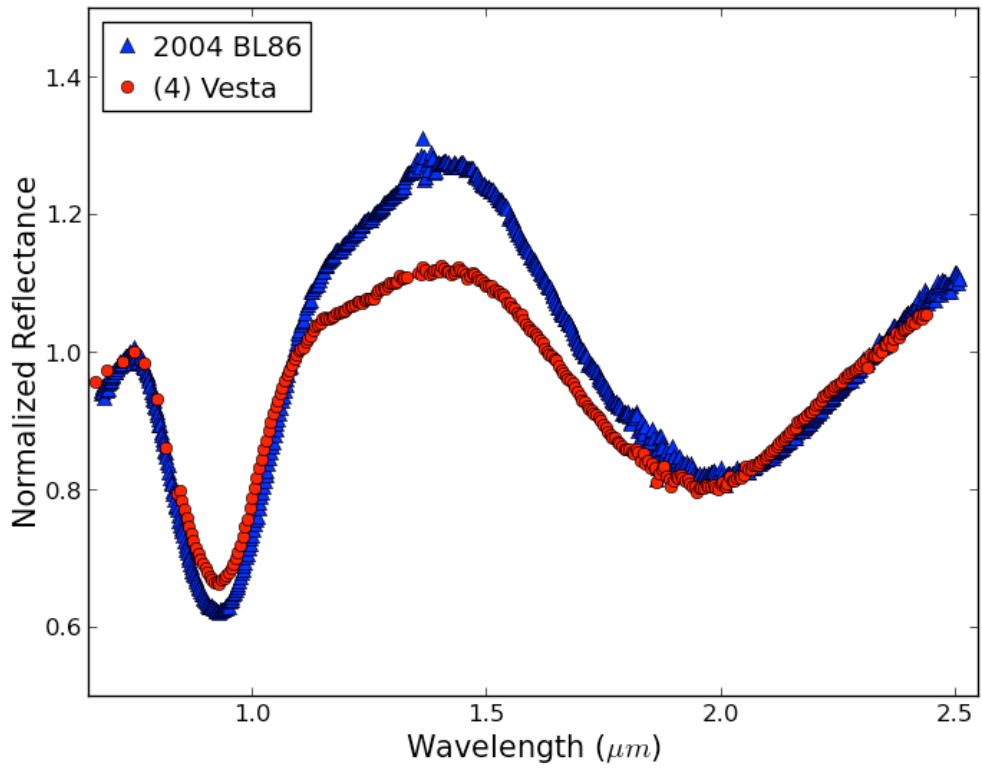